

\def\ifundefined#1{\expandafter\ifx\csname
#1\endcsname\relax}

\newcount\eqnumber \eqnumber=0
\def\beq{ \global\advance\eqnumber by 1 $$ }
\def\eeq{ \eqno(\the\eqnumber)$$ }
\def\label#1{\ifundefined{#1}
\expandafter\xdef\csname #1\endcsname{\the\eqnumber}
\else\message{label #1 already in use}\fi}
\def\(#1){(\csname #1\endcsname)}
\def\puteqno{\global\advance \eqnumber by 1 (\the\eqnumber)}

\newcount\refno \refno=0
\def\[#1]{\ifundefined{#1}\advance\refno by 1
\expandafter\xdef\csname #1\endcsname{\the\refno}
\fi[\csname #1\endcsname]}
\def\refis[#1]{\item{\csname #1\endcsname.}}

\def\sgn{ {\rm sgn}}
\def\half{ { 1 \over 2}}
\def\tr{{\rm tr}}


\baselineskip=18pt
\magnification=1200
\vskip1in
UR-1283

ER-40685-733

\vskip1in
\centerline{\bf Two Dimensional Quantum Chromodynamics on a Cylinder}
\vskip.5in
\centerline{S. Guruswamy and S.G. Rajeev}
 \centerline{\it Department of Physics and Astronomy}
 \centerline{\it University of Rochester}
 \centerline{\it Rochester, N.Y. 14627}
\vskip2in
\centerline {\bf ABSTRACT}

\noindent We study two dimensional Quantum
Chromodynamics  with massive quarks on  a
 cylinder  in a light--cone formalism. We eliminate the non--dynamical degrees
 of freedom and express the theory in terms of  the quark and Wilson loop
 variables. It is possible to perform this reduction without gauge fixing. The
 fermionic Fock  space can be defined independent of the gauge field  in this
 light--cone   formalism.
 \vfil\eject


                The problem of deriving the low energy properties of strong
 interactions from
 the lagrangian of Quantum Chromodynamics (QCD) remains as an important
 challenge to particle theorists. 't Hooft \[thooft] derived the meson spectrum
 in the large $N_c$ limit from  QCD in two dimensional Minkowski space.
 Recently, his  diagrammatic method has been reformulated in the language of
 operators\[meson] and the baryon\[2dbaryon] spectrum in the large $N_c$ limit
 has been derived as well.  In this model  there are no gluonic degrees of
 freedom; all the degrees of freedom in the  Yang--Mills  field can be  gauged
 away.

                If the space--time is not simply connected, there are some
Yang--Mills degrees
 of freedom that cannot be gauged away: those associated to parallel transport
 around non--contractible loops. A simple special case is Yang--Mills theory on
 a cylinder, which was solved by canonical methods in Ref. \[ymcyl]. In this
 case the only physical
 degree of freedom is the Wilson loop around the cylinder; Yang--Mills theory
 reduces to quantum mechanics on a group manifold. It is of interest to study
 Yang--Mills theory coupled to fermions on a cylinder, as it will provide a
 generalization of 't Hooft's model that contains some gluonic degrees of
 freedom. In this paper,  we will   reduce the action of Dirac--Yang--Mills
 theory on a cylinder to a form in which all but a finite number of degrees of
 freedom of the gauge have been eliminated. We believe that this form of the
 theory can  be solved in the large $N_c$ limit.

We will perform our analysis in a co--ordinate system different from the
 Cartesian $(x,t)$ co--ordinates
 used to solve pure Yang--Mills theory in Ref. \[ymcyl].

\noindent We define $u=t+|x|$ , $dt=du-\sgn(x) dx$.

\noindent The metric, ${ds}^2={dt}^2-{dx}^2={du}^2-2\sgn(x) du dx $ .
Thus the metric tensor, $${\eta}_{\mu\nu}=\left(\matrix{1&-\sgn(x)\cr-\sgn(x)&0
\cr}\right)\quad\hbox{,}\quad
        {\eta}^{\mu\nu}=\left(\matrix{0&-\sgn(x)\cr -\sgn(x)&-1}\right)$$
        $$\sqrt{-\eta}=1$$
The points $(u,x)$ and $(u,x+2L)$ are the same on the cylinder.

This system $(u,x)$  is a variant of the usual light--cone formalism. The
 disadvantage of the Cartesian co--ordinates in
 our case is that   the energy appears quadratically in the mass shell
 condition:
\beq
        p_0^2-p_1^2=m^2
\eeq
so that there are two values of energies, differing by a sign,   for each
 momentum. Then, the fermionic Fock space has to be defined  by filling the
 negative energy sea. Unfortunately, this can lead to trouble since the
 definition  of the Dirac sea can depend on the gauge field \[mickbook].
It is an old observation of Dirac \[dirac] ( revived  in the context of QCD in
 Ref.\[wilson]) that in light--cone co--ordinates, the analogue of energy
 appears linearly so that this problem does not occur.  If we were really to
use
 the conventional  light--cone co--ordinates, our method of eliminating the
 gauge  field  would
 not work. We will use $u$ as our evolution variable, so that the equal `time'
 surfaces are still light--cones.

The mass--shell condition  in our co--ordinate system is
\beq
        -2\sgn(x)p_x p_u-p_x^2=m^2
\eeq
so that $p_u$ has a unique solution:
\beq
        p_u=-\half[ {m^2\over p_x}+p_x]\sgn(x).
\eeq
If we impose $\sgn(x)p_x<0$ on the one--particle Hilbert space of fermions, we
 guarantee that the Fermionic energy is positive. The equal $u$ surface is a
 cone
 and  our condition says that the fermions move towards the future. For the
 moment we set aside this issue.

The Gamma matrices satisfying the Clifford algebra
$\bigl \{{\gamma}^\mu {,} {\gamma}^\nu \bigl \}= 2 {\eta}^{\mu \nu}$
are:
$${\gamma}^u={\gamma}^0+\sgn(x){\gamma}^1=\left(\matrix{0&1+\sgn(x)\cr
1-\sgn(x)&0\cr}\right)\quad\hbox{and}\quad
{\gamma}^1=\left(\matrix{0&1\cr-1&0\cr}\right)$$
Also,$${\gamma}^1{\gamma}^u=\left(\matrix{1-\sgn(x)&0\cr0&-1-\sgn(x)}\right)
\quad
\hbox{,}\quad{\gamma}^u{\gamma}^1=\left(\matrix{-1-\sgn(x)&0\cr0&1-\sgn(x)\cr}
\right)$$
With hindsight, in order to have only one component of the quark spinor to
evolve in time,we choose the quark spinor as:
                $$q={1\over{\sqrt2}}\left(\matrix{\chi\cr\psi}\right)\quad{,}
\quad  x >0$$
$$\quad={1\over{\sqrt2}}\left(\matrix{\psi\cr-\chi}\right)\quad {,}\quad x<0$$
On the cylinder,the points $(t,x)\sim(t,x+2L)$ i.e,$(u,x)\sim(u,x+2L)$ .
All fields are periodic with  period 2L:
$$q(u,x)=q(u,x+2L)$$
$$A_\mu (u,x)=A_\mu (u,x+2L)$$
We choose $ A_\mu$  to be anti-hermitian.

\noindent The  action of two dimensional QCD is,
\beq{\cal S}=\int du dx{\sqrt-\eta}\quad \Bigl [-{\bar q}i{\gamma}^\mu
({\partial}_\mu  +{A}_\mu)q-m{\bar q}q
+{1\over{4\alpha}}\tr {\cal F}^{\mu\nu}{\cal F}_{\mu\nu} \Bigl ]\eeq
In terms of the components of the quark spinor ( and introducing $E$ as an
 independent variable to get a first order action),
\beq\eqalign{{\cal S}=\int du dx\quad \Bigl [& -\psi^{\dag} i({\partial}_u
+A_u)
\psi-{\sgn(x)\over 2}\{\psi^{\dag} i({\partial}_1+A_1)\psi-\chi^{\dag}
 i({\partial}_1+A_1)\chi\}\cr&
-{m \sgn(x)\over2}(\psi^{\dag}\chi+\chi^{\dag}\psi)\cr&
-{1\over \alpha}\tr E\{{\partial}_uA_1-{\partial}_1A_u+[A_u,A_1]\}+
{1\over{2\alpha}}\tr{E}^2 \Bigl]\cr}$$
where  $$ E={\partial}_uA_1-{\partial}_1A_u+[A_u,A_1]\eeq
is the electric field.

Our strategy will be similar to that in Ref. \[ymcyl]: to eliminate the
 Yang--Mills field without imposing any gauge condition.

\noindent It will be useful to define  the variable {\it h} such that,
$${\partial {\it h}\over\partial x}+A_1{\it h}=0
\qquad {\rm with}\quad {\it h}(u,-L)=1.$$
Then ,$${\it h}(u,x)= P[e^{-\int_{-L}^{x}dy\quad A_1(u,y)}]$$
\noindent The Wilson loop is then given by  $ {\sl q}={\it h}^{-1}(u,L).$ In
 general, this $q$ is not equal to one; in fact it contains all the gauge
 invariant degrees of freedom of the gauge field co--ordinates. ( We will see
 that the boundary value of the electric field plays the role of a canonical
 conjugate to $q$).

\noindent Now,define new variables $\tilde{\psi}$ etc. by ,
$$\psi = {\it h} \tilde\psi\qquad
\chi = {\it h} \tilde\chi$$
$$A_u={\it h}{\tilde A_u}{\it h}^{-1}\qquad
E={\it h}\tilde E{\it h}^{-1}.$$

\noindent The old variables satisfied the boundary conditions:
        $${\chi}(L)={\psi}(-L)\qquad
{\psi}(L)=-{\chi}(-L)$$
$${\chi}(0^+)={\psi}(0^-)\qquad
{\psi}(0^+)={-\chi}(0^-)$$
 and
$$E(L)=E(-L).$$
In terms of the new field variables ,we see that :
$$\tilde \chi(L)={\sl q}{\tilde \psi(-L)} \qquad
\tilde \psi(L)=-{\sl q}{\tilde \chi(-L)}$$
$$\tilde E(L)={\sl q}{\tilde E(-L){\sl q}^{-1}}$$

\noindent We will define $E(-L)= e$.

\noindent After some calculations, the action can be written in terms of
the new field variables as :
\beq\eqalign{{\cal S}=& \int du dx \quad
\Bigl [-{\tilde \psi^{\dag}} i {\partial}_u{\tilde \psi} -{\sgn(x)\over 2}
({\tilde \psi^{\dag}}i{\partial}_1{\tilde \psi}
-{\tilde \chi^{\dag}}i{\partial}_1{\tilde \chi})
-{m \sgn(x)\over 2}({\tilde \psi^{\dag}}{\tilde \chi}+
{\tilde \chi^{\dag}}{\tilde \psi})\Bigl ]\cr &
+\int du dx \quad \Bigl [-{\tilde \psi^{\dag}}iA{\tilde \psi}
-{1\over \alpha}\tr\{({\partial}_1 {\tilde E})A\}\Bigl ]
+ {1\over {2\alpha}}\tr{\tilde E}^2 \Bigl ] \cr &
-{1\over{\alpha}}\int du \quad \tr({\sl q}^{-1}{\partial}_u{\sl q}e)\cr}\eeq

\noindent The action is written in the above form by using the identities
 $${\partial}_1A_u+[A_1,A_u]={\it h}({\partial}_1{\tilde A_u}){\it h}^{-1}$$
and
$${\it h}^{-1}({\partial}_uA_1){h}=
-{\partial}_1({\it h}^{-1}{\partial}_u{\it h}).$$
(Both of these can be seen using ${\partial}_1{\it h}=-A_1{\it h}$).

\noindent Also we have redefined,
$$ \tilde{A}_u+h^{-1}\partial_u h=A $$

 In this action, the variables $A$ and $\chi$ do not have
 derivatives with respect to the evolution variable $u$.
 Hence they are just Lagrange multipliers imposing some constraints
 and can be eliminated.
Varying the action written above w.r.t A, we obtain the constraint equation
( analogue of Gauss' law):
$${1\over\alpha}{\partial}_1{\tilde E}-:i{\tilde \psi}{\tilde \psi^{\dag}}:
\quad= 0$$
We can solve for $ {\tilde E(x)}$ :
$$\tilde E(x)= {\alpha}\int_{-L}^{x} dy \quad : i{\tilde \psi}
{\tilde \psi^{\dag}}(y): +\quad e$$
We still have  a constraint, coming from the periodicity condition
$E(-L)=E(L)$:
\beq e+{\alpha}\int_{-L}^{L} dy \quad :i {\tilde \psi}{\tilde \psi^{\dag}}(y):
\quad ={\sl q}e{\sl q}^{-1} \eeq
This generates a finite dimensional part of the gauge invariance that cannot be
 eliminated without running into Gribov ambiguities. At end we can recover this
part of the gauge group by imposing an equivariance condition on the
 wave--functions.

Also, we can eliminate ${\tilde \chi}$ from our action using :
\beq -i{\partial}_1{\tilde \chi}+m{\tilde \psi}=0 \eeq
This eliminates half the fermion degrees of freedom as in the usual light cone
 formalism \[meson].

We have thus reduced the action to:
\beq\eqalign{{\cal S}=& {1\over{2\alpha}}\int du dx \quad
\tr \Bigl [e+{\alpha}\int_{-L}^{x} dy \quad :i {\tilde \psi}
{\tilde \psi^{\dag}}(y):\Bigl ]^2
-{1\over\alpha}\int du \quad \tr({\sl q}^{-1}{\partial}_u{\sl q}e)\cr &
+\int du dx \quad \Bigl [-{\tilde \psi^{\dag}}i{\partial}_u{\tilde \psi}
+{\sgn(x)\over 2}{\tilde \psi^{\dag}}({\hat p}+{m^2 \over {\hat p}})
{\tilde \psi} \Bigl ] \cr}\eeq
where ${\hat p}=-i{\partial}_1$.

\noindent   Now we can read off the canonical commutation relations on
 an equal $u$ surface:
\beq
        [\tilde\psi^{\dag}(x),\tilde\psi(y)]_+=\delta(x-y)\quad
                         [\tr \lambda e,q]=\lambda q-q\lambda\eeq
 all others being zero. The hamiltonian is then
\beq\eqalign{ H=: & -{1\over{2\alpha}}\int  dx \quad
\tr \Bigl [e+\int_{-L}^{x} dy \quad :i{\tilde \psi} {\tilde \psi^{\dag}}(y):
\Bigl]^2  \cr & -\int  dx \quad \Bigl
 [{\sgn(x)\over 2}{\tilde \psi^{\dag}}({\hat p}+{m^2 \over {\hat p}})
{\tilde \psi} \Bigl ]: \cr}\eeq
It should be of interest to diagonalize this hamiltonian numerically using the
 techniques of Ref. \[paulibrodsky]. Alternately, one could `bosonize' this
 using the techniques of Ref. \[meson]. This would yield a theory of mesons,
 baryons and `glueballs'  that is semi--classical in the large $N_c$ limit.

\noindent We notice here that the current
 $j(y)= :i{\tilde \psi}{\tilde \psi^{\dag}}(y):$
is not periodic in 2L but,

$ {\tilde E}(x)=e+{\alpha}\int_{-L}^{x} dy \quad
 :i{\tilde \psi} {\tilde \psi^{\dag}}(y):$
obeys the periodicity condition:
$${\tilde E}(L)={\sl q} {\tilde E}(-L){\sl q}^{-1}$$ This is because the
 `electric field' operator has a geometric meaning as a covariant derivative on
 the associated bundle over the space of gauge fields modulo gauge
 transformations.

While this paper was in preparation, we received a recent paper \[lang] also
 dealing with two dimensional gauge theories. However, the approach is
 different, as they do not use a light cone formalism.

\eject

 { \bf REFERENCES}

\[thooft] G.'t Hooft,Nucl. Phys. B75, 461 (1974)

        and E.Witten, Nucl. Phys. B 160, 57 (1979).

\[meson] S.G.Rajeev ,`` Two Dimensional Meson Theory '',1991 {\it Summer School

 in High Energy Physics and Cosmology},edited by E.Gava et al, World Scientific
(1992).

\[2dbaryon] P.F.Bedaque, I.Horvath, S.G.Rajeev,`` Two Dimensional Baryons in
the
    Large

    N Limit '' (to be published in Mod. Phys. Lett.).

\[ymcyl] S.G.Rajeev, Phys. Lett. 209B:53 (1988).

\[mickbook] J.Mickelsson ,{\it Current Algebras and Groups} , Plenum Monographs
   in

    Nonlinear Physics (1989).

\[dirac] P.A.M.Dirac, Rev. Mod. Phys. 21, 392 (1949).

\[wilson] R.J.Perry, A.Harindranath and K.G.Wilson, Phys. Rev. Lett. 65, 2959
          (1990).

\[paulibrodsky] K.Hornbostel,S.Brodsky,H.Pauli, Phys. Rev.,41, 3814 (1990)

\[lang] G. Semenoff and E. Langmann, University of British Columbia preprint (
 1992).

\end